\documentclass[sigplan,screen,nonacm]{acmart}

\AtBeginDocument{%
  }

\usepackage{epstopdf}
\usepackage{multirow}
\usepackage{algorithm}
\usepackage{algorithmic}
\usepackage{tikz}
\usepackage{tcolorbox}
\usepackage{makecell}
\usepackage{subfigure}

\begin{document}

\title{SAIH: A Scalable Evaluation Methodology for Understanding AI Performance Trend on HPC Systems}

\author{Jiangsu Du, Dongsheng Li, Yingpeng Wen, Jiazhi Jiang, Dan Huang, Xiangke Liao, and Yutong Lu$^*$ \\
School of Computer Science and Engineering, Sun Yat-sen University, Guangzhou, 510006, China \\
dujs@mail2.sysu.edu.cn, luyutong@mail.sysu.edu.cn
}
\thanks{* Corresponding Author.}
\thanks{The manuscript has been submitted to the Journal of Computer Science and Technology and gets the minor revision decision in Dec 2021.}
\renewcommand{\shortauthors}{Du et al.}

\begin{abstract}
Novel artificial intelligence (AI) technology has expedited various scientific research, e.g., cosmology, physics and bioinformatics, inevitably becoming a significant category of workload on high performance computing (HPC) systems. Existing AI benchmarks tend to customize well-recognized AI applications, so as to evaluate the AI performance of HPC systems under predefined problem size, in terms of datasets and AI models.
However, driven by novel AI technology, most of AI applications are evolving fast on models and datasets to achieve higher accuracy and be applicable to more scenarios. 
Due to lack of scalability on the problem size, static AI benchmarks might be under competent to help understand the performance trend of evolving AI applications on HPC systems, in particular, the scientific AI applications on large-scale systems.
In this paper, we propose a scalable evaluation methodology (SAIH) for analyzing the AI performance trend of HPC systems with scaling the problem sizes of customized AI applications.
To enable scalability, SAIH builds a set of novel mechanisms for augmenting problem sizes.
As the data and model constantly scale, we can investigate the trend and range of AI performance on HPC systems, and further diagnose system bottlenecks.
To verify our methodology, we augment a cosmological AI application to evaluate a real HPC system equipped with GPUs as a case study of SAIH. 
With data and model augment, SAIH can progressively evaluate the AI performance trend of HPC system, e.g. increasing from 5.2\% to 59.6\% of the peak theoretical hardware performance.
The evaluation results are analyzed and summarized into insight findings on performance issues.
For instance, we find that the AI application constantly consumes the I/O bandwidth of shared parallel file system during its iteratively training model. 
If I/O contention exists, the shared parallel file system might become a bottleneck.
\end{abstract}




\maketitle

\section{Introduction}

In recent years, AI, especially its deep learning subset, becomes one of the key trends in HPC.
Emerging novel AI applications have expedited various scientific discoveries, such as in cosmology~\cite{ravanbakhsh2016estimating, mathuriya2018cosmoflow}, physics~\cite{kurth2017deep, patton2018167}, cancer diagnosis~\cite{balaprakash2019scalable, wozniak2018candle} and etc. 
Besides, scientists significantly improve domain results with AI technology on HPC systems over their traditional competitors~\cite{kurth2018exascale}.

To support AI applications, emerging HPC systems are designed towards exascale computing capability with better AI performance in consideration.
For example, Fugaku in Japan exhibits $2.0\times$ mixed-precision exaflops with ARM based many-core CPUs and Summit at ORNL has $1.4\times$ mixed-precision exaflops with NVIDIA GPU.
While the theoretical peak performances of these HPC systems are appealing, their practical performance on supporting AI training (\emph{HPC-AI performance}) are still under investigation.
In terms of traditional HPC benchmarks, e.g., HPL and HPL-AI~\cite{hpl}, they parallelly solve a common task in HPC domain, the linear equation $Ax = b$, which is rarely adopted in AI applications and thus lacks convincible metrics on AI performance.
Compared to traditional HPC simulations, scientific AI applications are much more complex.
HPC simulations typically consist of distinct execution phases, such as computation, I/O and collective communication.
In contrast, the execution phases of AI applications are pipelined and encapsulated by higher abstractions and longer parallelism hierarchy, that include co-designed programming models, e.g., CUDA, and distributed AI frameworks, e.g., Tensorflow and PyTorch. 
Such abstraction complexity leads to doubt on the significance of profiling the AI performance by traditional HPC benchmarks.

Moving to existing AI benchmarks on HPC systems, e.g., CORAL-2~\cite{coral2} and MLPerf~\cite{MattsonCDCMPTWB20}, they are endeavoring to evaluate either the accuracy or the performance (flops) of AI models in the hotspot research domains by adopting representative AI applications with static data and models. This static methodology can reflect a fixed performance relation between a specific application and HPC system, rather than a range of performance relations.
Although MLPerf recently focuses on HPC systems, and adds scientific AI applications, e.g., Cosmoflow~\cite{mathuriya2018cosmoflow} and DeepCAM~\cite{kurth2018exascale}, into its benchmark suit, the main idea still follows its previous version, in which both models and datasets are static.

Driven by novel AI technologies, most of AI applications are evolving fast on its problem configurations including both models and datasets, for the purpose of achieving higher accuracy and being applicable to more scenarios. As proposed in NVIDIA GTC 2021~\cite{gtc2021}, AI model sizes are growing exponentially, on a pace of doubling every two and a half months. Due to lack of scalability on the problem size, existing AI benchmarks might be incompetent to understand the performance trend of evolving AI applications on HPC systems, in particular, the scientific AI applications on large-scale systems. 
For instance, in our evaluation, we observe that the aggregate flops of a parallel run of Cosmoflow (3D CNN) is significantly increased (from 5.2\% to 59.6\% of peak theoretical performance) with the scaling datasets and AI models.
It reflects that a kind of AI workloads vary vastly in different configurations and the execution of a static application can only provide a very partial understanding.

To understand AI performance trend on HPC systems in a more comprehensive way, we propose the scalable evaluation methodology (SAIH).
Like the successful HPL LINPACK benchmark on achieving the data and computation scalability by adjusting the size of $Ax=b$, SAIH builds a set of novel mechanisms to satisfy the requirements of data and computation scalability. Additionally, SAIH takes into account the scientific significance of AI workloads by selecting and building representative study cases from domain-specific scientific AI applications.
Specifically, the contributions of SAIH are as follows.
\begin{itemize}
    \item We propose SAIH with scientific significance to evaluate and understand the AI performance range of HPC systems, and it is with both data and model scalability to cover various problem sizes.
	\item We design a novel strategy for model scalability. By creatively incorporating network architecture search (NAS), the strategy extends original AI model to more accurate and complex models with scaling computation demands. 
	\item We implement a cosmological AI application as a prototype and rebuild it to a SAIH instance with data and computation scalability. We apply this instance to evaluate a real HPC system as a case study. 
	\item We summarize the performance achievement and qualitative evaluation metrics to illustrate that the SAIH instance can profile the AI performance range of HPC system on a specific scientific domain as well as reveal evaluation findings about the potential performance issues.
\end{itemize}

The rest of this paper is organized as follows. We present a survey on scientific AI applications and existing AI benchmarks along with the related work in Section~\ref{bg_rw}. A comprehensive methodology on building scalable AI evaluation is presented in Section~\ref{method}. In Section~\ref{case}, we present a case study on building a SAIH instance based on a cosmological AI application. Section \ref{eval} presents the evaluation and analysis of the cosmological SAIH instance on a real HPC system.
Finally, we conclude this work in Section~\ref{conclusion}.

\section{Background and Related Work}
\label{bg_rw}
\subsection{Scientific AI Applications}

In recent years, scientific AI applications are emerging in various domains. 
In particle physics, Kurth~\cite{kurth2017deep} firstly attempted to deploy particle image classification on many-core HPC systems and achieved peta-flops performance. 
Patton and et al.~\cite{patton2018167} adopt neural network search to find an optimal model for improving the understanding of the electron-beam-matter interactions and realtime image-based feedback, which enables a huge step beyond human capacity towards nano-fabricating materials automatically. 
In the domain of astronomy, cosmologists~\cite{ravanbakhsh2016estimating, mathuriya2018cosmoflow} take advantage of CNN models to estimate the universal states with higher accuracy than traditional methods. 
The universal parameters are key factors that determine the evolution of the whole universe and classification and discovery of astrophysical objects.
Medical imaging analysis is the science of solving clinical problems by analyzing images generated in clinical practice. 
Deep learning techniques are applied in computer aided diagnosis by analyzing the signal data from CT, MRI, DR, and etc, including image segmentation~\cite{zhang2011bayesian}, detection and classification of abnormality~\cite{ balaprakash2019scalable, wozniak2018candle}. 
In bioinformatics, RNN based deep learning techniques are widely used to detect and recognize genomic patterns~\cite{shen2018recurrent,  trabelsi2019comprehensive, lyu2017long}.
In climate changing and weather analysis, CNN and RNN models are used to detect the areas where the abnormal climate changes~\cite{kurth2018exascale,karpatne2017big}.

\vspace{-5pt}
\subsection{AI Benchmarks}
\vspace{-5pt}
AI based techniques are continuously driving various application scenarios intelligent, which can be deployed on diverse computing platforms, from large-scale HPC systems to tiny mobile devices.
To satisfy the demands of evaluating the AI performance, scientists and engineers have released a number of AI benchmarks covering different application scenarios and platforms.

MLPerf~\cite{MattsonCDCMPTWB20} is an AI benchmark suite targeting six AI application scenarios, including recommendation, speech recognition, reinforcement learning, image classification, object detection, and translation.
DeepBench~\cite{Deepbench}, released by Baidu, is a micro benchmark set that evaluates basic operations involved in training deep neural networks, including dense matrix multiplies, convolutions, recurrent layers, and all-reduce. 
This benchmark lacks the component-level and application-level evaluation cases.
AI Matrix~\cite{zhang2019ai}, released by Alibaba Group, aims to satisfy the needs of fully characterizing the deep learning workloads in Alibaba's e-commerce environment, including the tasks in computer vision, recommendation and language processing.
HPL-AI Mixed Precision Benchmark~\cite{haidar2018harnessing}, released by University of Tennessee, is opting for low-precision (likely 16-bit) accuracy for LU (lower-upper) factorization, and a sophisticated iteration to recover the accuracy lost in factorization.
Deep500~\cite{Deep500} is a modular benchmark infrastructure for high-performance computing deep learning. It aims at evaluating different framework implementations and different levels of operators. 
However, it only evaluates common image classification scenario on ImageNet~\cite{imagenet} dataset, rather than typical scientific scenarios.
TBD Suite~\cite{zhu2018tbd}, developed by University of Toronto, is an end-to-end benchmark suite for neural work training. Typically, this work currently covers six major application domains and eight different state-of-the-art models, e.g., image classification, speech recognition and etc. 
The above AI benchmarks provide evaluations either for classic application scenarios or for the computing operations in deep nerual network (DNN) models on data centers and mobile devices.
However, they have not covered the scientific AI applications and HPC systems, 
the configurations of which are significantly distinct from evaluation cases of existing AI benchmark in both datasets and DNN models.

Recently, a number of well-recognized scientific AI applications are integrated into HPC benchmark suite for the purpose of evaluating the performance of AI application on HPC systems.
For example, CORAL-2 benchmarks~\cite{coral2} cover not only the traditional micro benchmarks, HPC simulations and analysis, but also emerging AI applications, which include the common operations in CNN and RNN as well as the CANDLE benchmark~\cite{wozniak2018candle} for cancer diagnosis.
In 2020, two scientific applications, cosmoflow~\cite{mathuriya2018cosmoflow} and DeepCAM~\cite{kurth2018exascale} with static models and datasets are included into MLPerf, named as MLPerf-HPC~\cite{mlperfhpc}, for evaluating the AI performance of large HPC systems. 
In general, HPC-AI500~\cite{hpcai500} follows a similar idea as MLPerf-HPC. Moreover, in order to assure a fair ranking, it presents a new metric named Valid FLOPS, which imposes a penalty on failing to achieve a target training quality.
While CORAL-2 benchmarks, MLPerf-HPC, and HPC-AI500 are capable of distributedly evaluating scientific AI applications with static scientific AI models and datasets on HPC systems, they lack the capability of scaling up problem sizes, which is critical to evaluating the potential AI performance of large-scale HPC systems.

\section{Methodology}
\label{method}

\begin{figure}[!t]
  \centering
    \includegraphics[width=3in]{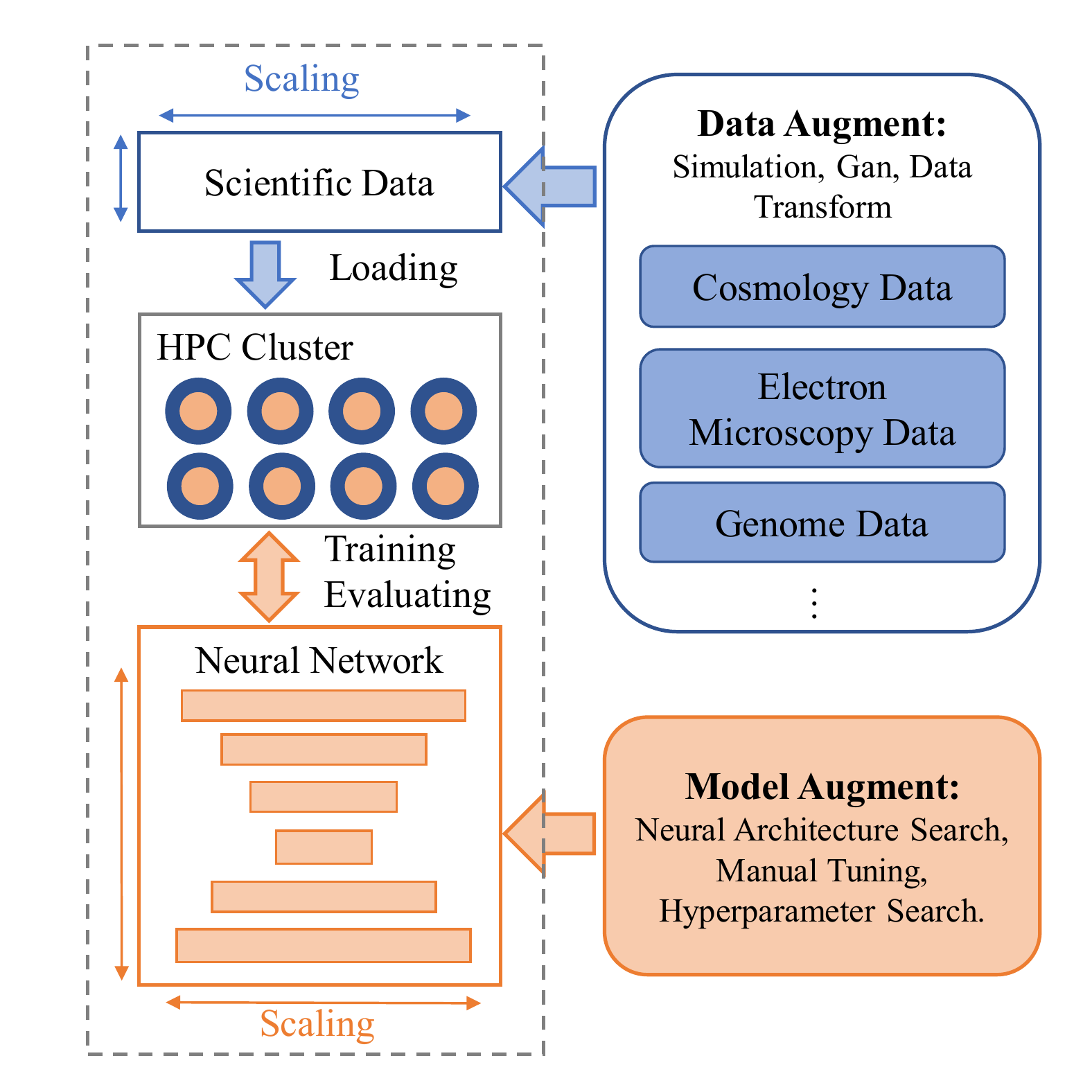}
  \caption{SAIH overview.}
  \label{fig:design}
\end{figure}

In this section, we introduce the methodology of SAIH. 
While scientific AI applications run on large-scale HPC systems in data-parallel fashion via sophisticate AI frameworks, such as MPI, Tensorflow and PyTorch, the problem sizes are typically limited by the static training datasets and AI models. 
Thus, we intend to scale up problem sizes as well as the demands on memory and computation resources by integrating the methods of data augment and AI model augment.
Our augment methods take the scientific meaning into account, and higher resolution or higher accuracy can be achieved than original applications.
As shown in Fig.~\ref{fig:design}, SAIH provides a set of augment methods for transforming scientific AI application into candidate benchmarks with the scalability on problem sizes, and then evaluating the AI performance of HPC systems. 

\subsection{Data Augment} 

Data augment method in SAIH is designed for achieving both data scalability and keeping scientific meaning.
Thus, simulation can be a potential method since it is naturally supported by many scientific applications.
Besides, emerging generative adversarial network (GAN) is also a potential method.

\textbf{Simulation:} 
Many scientific applications simulate natural phenomena and the runtime states of large research facilities. 
To discover the scientific insights, the outputs of these simulations are the raw data that need to be further processed by analysis techniques, such as visualization tools and AI based analysis. 
To efficiently generate hundreds and thousands of data samples, we manage the initial parameters by a centralized parameter server to ensure parameters between different simulations randomly and uniquely.
The detailed procedure is in Algorithm~\ref{alg1}.

\begin{minipage}{0.4\textwidth}%
	\begin{algorithm}[H] 
		\caption{Concurrently generate data by simulation} 
		\label{alg1} 
		\begin{algorithmic}[1] 
			\STATE Start MPI based program with total N processes, one for parameter server and each of the ($N-1$) processes run $M$ instances of the simulation.
			\STATE Initialize MPI.
			\IF{$ rank = 0$}
			\STATE Initialize a centralized parameter server.
			\STATE Randomly generate initial parameters for simulations.
			\ELSE
			\STATE Wait for parameter server ready and prepare $M\times(N-1)$ parameters.
			\WHILE{$m < M$}
			\STATE Load parameters from server.
			\STATE Initialize an instance of simulation.
			\STATE Perform simulation.
			\STATE Save generated data into filesystem.
			\STATE $m=m+1$
			\ENDWHILE
			\ENDIF
		\end{algorithmic}
	\end{algorithm}
\end{minipage}

\begin{figure}[H]
	\centering
	\includegraphics[width=3.2in]{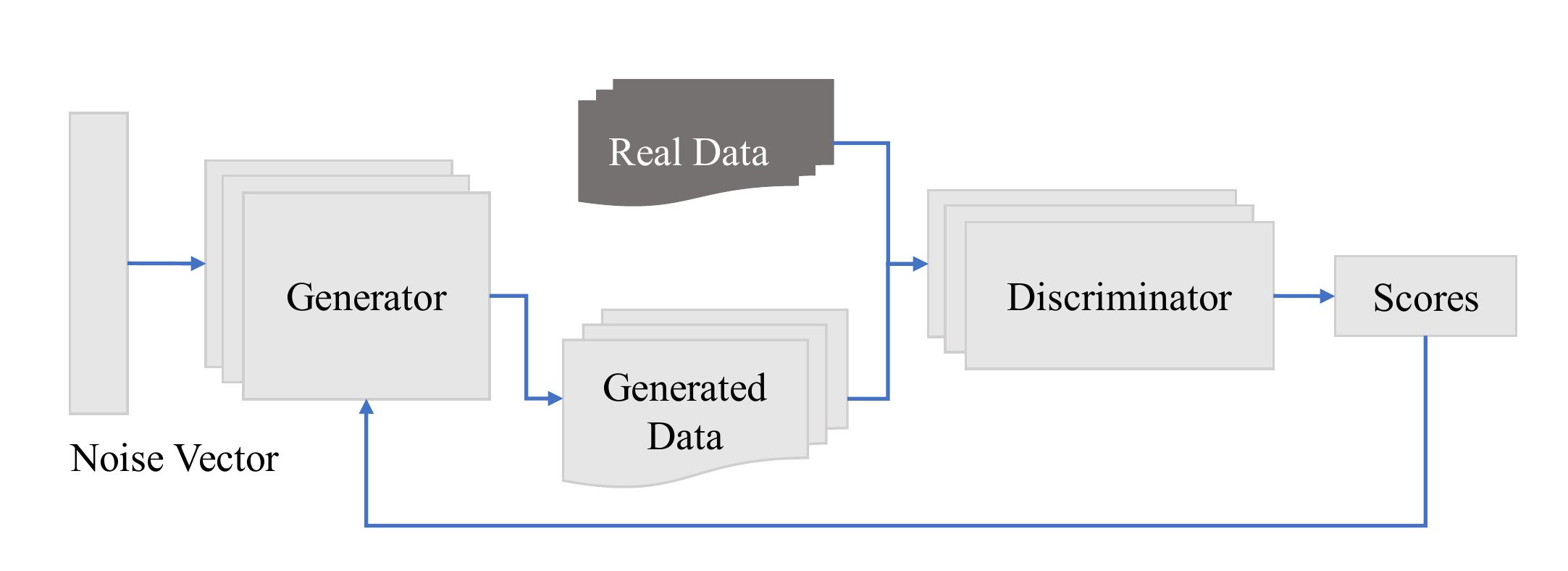}
	\caption{GAN for data augment.}
	\label{fig:gan}
	\vspace{-10pt}
\end{figure}

 \textbf{Generative adversarial networks}: 
GAN is another method of generating training data with demanded structures and properties for scientific applications.
We can use GAN to explore the space of possible data, tuning the generated data to have specific target properties.
In particular, GAN (CycleGAN) is used to augment CT images and improve generalizability in CT segmentation tasks~\cite{ganct}. 
GAN is also adopted to augment 3D MRI data for medical image segmentation~\cite{ganmri}.
Besides augmenting biomedical datasets, GAN can be also used to generate multi-sensor data for the aerial object detection and semantic segmentation on visual data, such as 3D Lidar reconstruction using ISPRS and DOTA datasets~\cite{ganaerial}.

As shown in Fig.~\ref{fig:gan}, GAN has a pair of components competing with each other, where the generator model is responsible for generating new synthetic data (e.g., DNA sequence in Genome).
The discriminator model calculates the similarity score between the generated data and the real data. 
Concretely, the GAN models are different case by case.
As iteratively training the models, the accuracies of the two models are improved. 
With the trained generator by GAN, we can concurrently run it to generate a large amount of training data.

\subsection{Model Augment} 

\begin{figure}[H]
 \centering
 \includegraphics[width=3.3in]{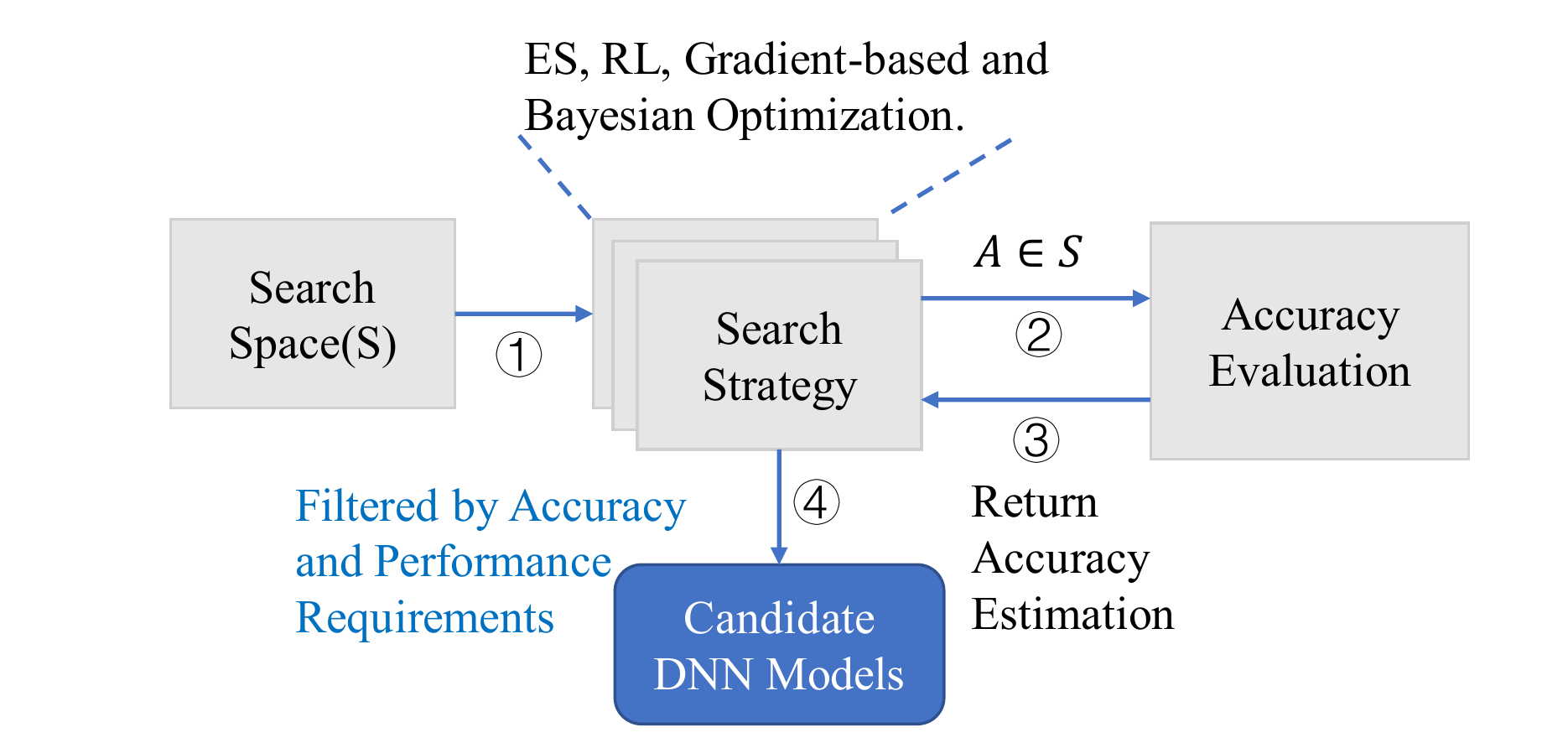}
 \caption{Network Architecture Search in SAIH.}
 \label{fig:nas}
 \vspace{-10pt}
\end{figure}

SAIH can scale its computation demand by model augment, which in deep learning is mainly either by AutoML~\cite{jin2019auto} based techniques such as NAS, or manually adjusted by AI experts.  
With these two kinds of model augment methods, generated candidate models can satisfy both scalability and accuracy requirements.
Although manually tuning is easier, it is time-consuming and requires a lot of expert knowledge.
In SAIH, we adopt NAS as the fundamental of our model augment method and a new component is inserted into NAS for selecting out the qualified models that has customized computation demand.

As shown in Fig.~\ref{fig:nas},
NAS generally consists of three main components: search space, search strategy, and accuracy evaluation.
The search space contains the architectures that can be selected. 
Users can set hyperparameters to restrict the searching properties of architectures for the purpose of reducing the size of the search space and simplify the search.
The search strategy specifies how to traverse in the search space, and is often exponential complexity, which motivates experts to design novel search optimizations or heuristic algorithms to find the optimal or near-optimal DNN models. 
The widely investigated optimizations include evolution strategy (ES), reinforce learning (RL), Bayesian optimization and etc. 
The accuracy evaluation is estimating the performance of model architecture, e.g., by performing a standard training and validation of the architecture on data.
When the estimation is evaluated and returned back, the search strategy will decide the next searching architecture, or accept the model based on the accuracy.

\vspace{-15pt}
\begin{align} \label{eq:1}
    Computation=nAddMul \times flopAddMul \times FB.
\end{align}

In order to generate models with a demanded computation cost, e.g., FLOPs, we design a novel component, model filter, to filter models with a demanded range of computation costs.
The model filter is integrated as a supplement component of the search strategy, enabling that SAIH explores models with a specific range of computation costs. 
Due to the long time of NAS, e.g., searching 20,000 neural networks across 500 P100 GPUs over 4 days~\cite{liu2018progressive}, in SAIH we tend to place the model filter before searching strategy, preparing candidate DNN models with scaled computation costs before training and evaluating.
The model filter estimates the theoretical computing cost of candidate model as the filtering basis.
We aggregate the number of multiply-accumulate operations to represent the computation cost.
It is straightforward to count the number of multiply-accumulate operations in the forward pass, while it is not for the backward pass, that is derived by the forward pass.
Here we adopt Equation~\ref{eq:1}~\cite{flopsmeasure} for estimating the computation cost of both forward and backward pass, where $numAddMul$ denotes the total number of add-multiply operations in the forward pass of a model, $flopAddMul$ is the number of FLOPs per add-multiply, and $FB$ is a constant of 3 for calculating both forward and backward pass.

\subsection{Performance and Qualitative Evaluation Summary}
\begin{table}[H]
	\footnotesize
	\centering
	\caption{Performance and Qualitative Evaluation Metrics for SAIH Case Study.}
	\label{tab:metrics}		
		\begin{tabular}{|m{2.3cm}<{\centering}|m{5.6cm}|}  
			\hline
			Metrics & Value \\
			\hline 
			Domain & HPC scientific domains, e.g., particle physics, astronomy, biomedical, climate, bioinformatics \\ 
			\hline
			Data augment & Simulation, GAN and data transformation \\
			\hline
			Model augment & Manual tuning, hyperparameter search and NAS  \\
			\hline
			DNN model & 2D CNN, 3D CNN, RNN, Transformer, etc.\\
			\hline
			Data format & 1D genome, 2D image, 3D particle, 3D tomography, etc. \\
			\hline
			AI framework& Tensorflow, PyTorch, Caffe, etc. \\
				\hline
			Dataset size & Dataset size range in weak scaling. \\
			\hline
			HPC system & The hardware configurations of an HPC system on which SAIH is performed. \\
			\hline
			FLOPs of AI model & Computation demand range for the candidate AI models. \\
			\hline
			flops (\% of theoretical flops) & The aggregate flops that is achieved by performing SAIH instance on an HPC system. \\
			\hline
			Singe GPU \newline performance & The single GPU performance range on data/node scaling. \\
			\hline
			Speedup on node scaling & The speedup range on strong scaling evaluation. \\
			\hline
			Arithmetic Intensity & The range of arithmetic intensity of AI models. \\
			\hline
			Accuracy/Loss & The range of accuracy/loss of AI models with data/node scaling. \\
			\hline
		\end{tabular}
\vspace{-5pt}
\end{table}

AI applications have been emerging from various scientific domains. 
These applications have two main components: data and model.
The data format and model architecture vary largely in different AI applications.
When we apply SAIH to a domain case, the evaluation can not only present the AI performance on HPC systems, but also the performance bottlenecks introduced by the characteristics of AI applications and frameworks, in terms of data movement, scalability and etc.
Thus, as shown in Table~\ref{tab:metrics}, we need to set a standard or metrics to summarize the evaluation results of SAIH instances.
The metrics table typically includes three main sections, 1) the characteristics of SAIH study case, 2) AI performance in terms of flops range, speedup, etc., and 3) other profiling results.

\section{Case Study}
\label{case}
In this section, we orchestrate a cosmological AI application, cosmoflow~\cite{mathuriya2018cosmoflow}, as a case study of SAIH, named as SAIH-cosmo.
Then, we apply SAIH-cosmo to an HPC system for understanding the relative AI performance trend.
We observe some findings stemming from the convergence of training model, performance bottlenecks on the software stack and hardware configurations and tuning. Such insights can guide the HPC systems for tuning better AI performance in the future.

\subsection{Background}
\subsubsection{Cosmoflow} 
In the domain of cosmology, it is of great significance to determine cosmological states, that can describe the evolution of the whole universe and the distribution features of matter and energy in a particular space.
Enabled by AI technology, the cosmoflow project aims at providing a faster, cheaper, and more accurate data processing workflow for cosmology research.
It employs 3D CNN to learn the mapping between the distribution of matters within a defined space and three cosmological states, by training the model on a large amount of simulated data.
With the trained model, cosmological states of a space can be inferred faster and more accurately, compared with traditional statistical methods.

\subsubsection{HPC System Configuration}
The system is a heterogeneous HPC cluster. 
Each node of the cluster is equipped with 2 Intel(R) Xeon(R) Gold 6132 CPUs (14 cores) operating at 256 GB memory and 4 Tesla V100 SXM2 GPUs.
Each Tesla V100 GPU has 16 GB HBM2 and can provide up to 7.8 Tflops double-precision performance, 15.7 Tflops single-precision performance and 125 Tflops half-precision performance.
Tesla V100 GPUs are interconnected by NVLink within the node and the nodes of the cluster are interconnected by InfiniBand.
The file system is Lustre for the persistence of training data.

\subsection{Data Augment and Preprocessing}

\begin{figure*}[!t]
	\centering
	\includegraphics[width=6.4in]{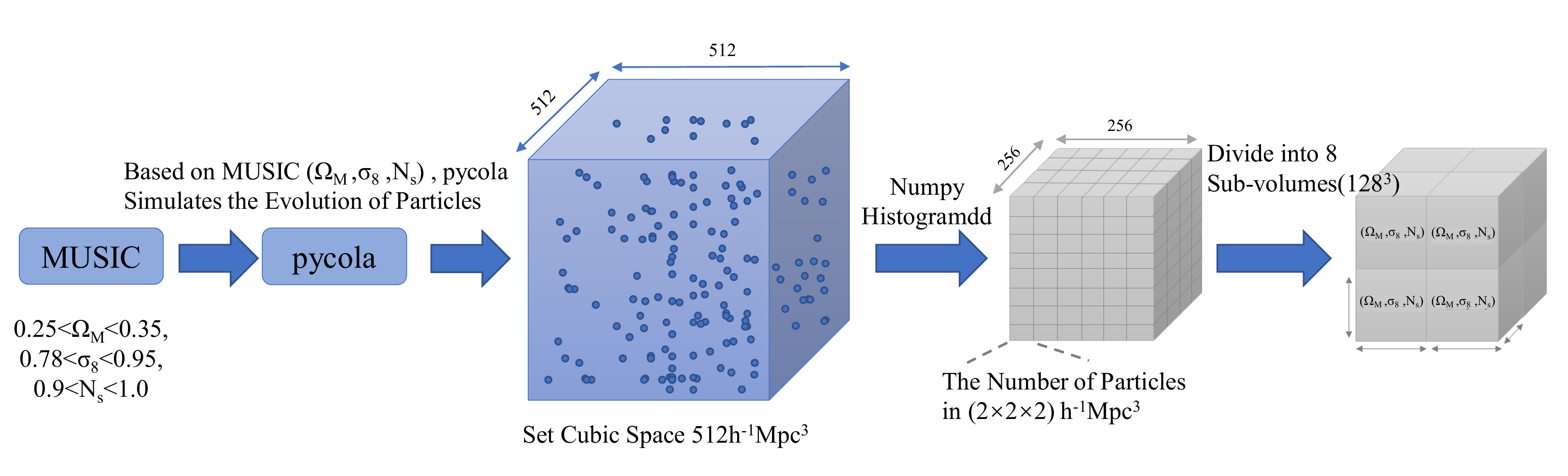}
		\vspace{-10pt}
	\caption{Data augment and preprocessing of the cosmological data.}
	\label{fig:dataset}
	\vspace{-10pt}
\end{figure*}

We use the simulation method for SAIH-cosmo, which adopts a widely-accepted simulator of generating cosmological data, e.g., adopted by cosmoflow~\cite{mathuriya2018cosmoflow}.
Different from the static datasets in cosmoflow of MLPerf HPC, we produce the dataset of SAIH-cosmo by modifying its simulation and make it scalable on dataset size, avoiding non-trivial data collecting, downloading and preprocessing.
The data is produced by N-body simulation, pycola~\cite{TassevPycola15}, of which the initial parameters are randomly generated by MUSIC~\cite{HahnMusic2011}.

In our case study, MUSIC randomly initializes matter distribution based on three parameters, consisting of $\Omega_M$, $\sigma_8$ and $N_s$, before the simulation runs.
There parameters can describe the cosmological state.
The three parameters are also kept as the label for its corresponding simulation output.
In detail, $\Omega_M$ is the proportion of matter in the universe.
We assume a flat geometry for universe, i.e., the sum of the contributions of matter and dark energy to the energy density of the universe $\Omega_M +\Omega_\lambda = 1$. 
$\Omega_\lambda$ is the proportion of the dark energy.
$\sigma_8$ is the amplitude of mass fluctuations in the universe at a distance scale of $8 Mpc/h$.
$N_s$ is the scalar spectral index for the spatial curvature of a comoving slice of the space-time.
The simulation, pycola, implements the comoving lagrangian acceleration (COLA) method in the temporal and spatial domains for the N-body simulations.
It simulates the evolution of physical particles from the MUSIC initial conditions.
The simulation result describes the position information of a predefined three dimensional spaces, such as $256^3$, $512^3$, and $1024^3$, which can be adjusted in initial condition of MUSIC. 
Essentially, the three cosmological states are set in the specific ranges, in which MUSIC can randomly pick values based on the uniform distribution.

We set the simulation cube $512^3$ relevant to $512h^{-1}Mpc^3$ cubic space as the leftmost cube of Fig.~\ref{fig:dataset} and the three cosmological states are randomly generated from $ 0.25<\Omega_M<0.35 $, $ 0.78<\sigma_8<0.95 $, and $ 0.9< N_s<1.0$.
MUSIC and pycola run on CPUs and consume massive resources, in particular memory.
Based on our evaluation, the memory requirement for an instance of the simulation with $512^3$ space size requires 17 GB memory for MUSIC and 20 GB for pycola.
For a single-node execution, the concurrency of MUSIC and pycola is constrained by memory capacity.
Since MUSIC and pycola run in series rather than concurrently, we can at most concurrently perform 12 simulations on each node.
Further, as in Algorithm~\ref{alg1}, we employ multi-node execution to further accelerate data augment.
As each simulation is independent, three cosmological states are randomly sampled within the predefined ranges via a third party math library, GNU Scientific Library.

After raw data is collected from pycola simulation, the data preprocessing is then performed to transform the raw data to training data that can be directly fed into 3D CNN model.
The data is transformed to volumetric form where a 3D histogram of $d^3$ voxels represents the normalized density of the dark matter for each cube.
As shown in Fig.~\ref{fig:dataset}, the resolution of voxel is $2\times2\times2 h^{-1}Mpc^3$, and the $512h^{-1}Mpc^3$ cube is transformed to $256^3$ voxel volumes. 
Then, each volume is split into 8 sub-volumes labeled with the same cosmological states as in the rightmost cube of Fig.~\ref{fig:dataset}.
Each sub-volume is 16 MB and stored in the HDF5 data format.
For this case study, we perform 12,632 simulations and create 101,088 sub-volumes as the total training dataset, taking around 1.6 TB storage.
In SAIH-cosmo, we extract a portion of the dataset for weak and strong scaling evaluations.

\subsection{3D CNN Model Augment}

\begin{figure}[H]
    \centering
 	\includegraphics[width=3.3in]{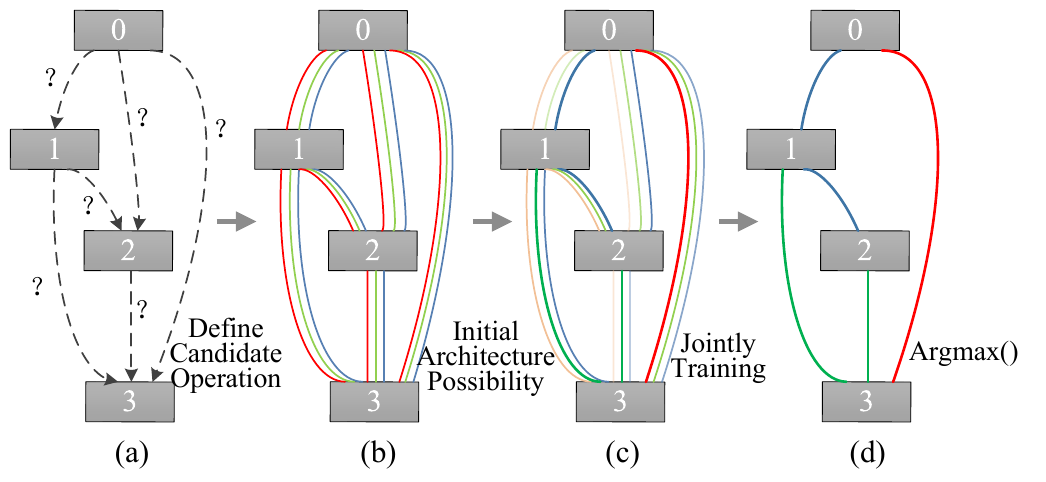}
 	\caption{DARTS in SAIH-cosmo.(a) Assign feature maps.(b) Define candidate operations for each edge. (c) Jointly learn the potential architectures and the network weights. (d) Select the final architecture.}
 	\label{fig:darts}
 	\vspace{-15pt}
\end{figure}

Here we take advantage of both manual design method and automated method.
For the manual design method,
we adopt the 3D CNN model proposed by Mathuriya et al.~\cite{mathuriya2018cosmoflow} as the smallest model with the least computation cost.
For the automated method, we customize NAS to prepare another two CNN models for SAIH-cosmo with incremental computation costs and model accuracies.
In details, we generate them by our proposed filter-based model augment method, which in this case adopts DARTS~\cite{NASLiuSY19} as its internal NAS strategy.
DARTS is a cell-based NAS technique with gradient descent search strategy.
The gradient descent search strategy is the 
non-differentiable approach and it is orders of magnitude faster than other non-differentiable approaches.
With the V100 GPU, DARTS only takes about 2 GPU days for exploring the medium model in our experiments, while other techniques require thousands of GPU days theoretically.
DARTS is a cell-based NAS technique, which searches the architecture of a cell and connects multiple identical cells as the entire model.
As shown in Fig.~\ref{fig:darts}(a), DARTS initially assigns a number of feature maps (nodes) in a cell, while edges between nodes are unknown.
Then, users define candidate operations for each edge based on observations from previous models.
Since each edge normally has only one operation, DARTS introduces the architecture possibility for each operation in each edge and determines the operation in each edge by gradient descent.
Training DARTS on a small dataset is adequate, which further accelerates the searching process. 
As shown in Fig.~\ref{fig:darts}(c), in each iteration, DARTS jointly learns the potential architectures and the network weights, which have the same target that improves the prediction accuracy.
In this way, the model constantly converges to the architecture with a higher accuracy.
Finally, DARTS selects the most likely operation of each edge to determine the final architecture.

Our 3D convolutional cell consists of $N=7$ nodes, and operation candidates include $3\times3$ and $5\times5$ 3D separable convolutions, $3\times3$ and $5\times5$ 3D dilated separable convolutions, $3\times3$ 3D max pooling, $3\times3$ average pooling, identity, and zero.
For each convolution operation, a batch-norm layer and the Leakey Relu activation function are followed behind based on observations from the smallest model. 
Since a single sample in our dataset is larger than experiments in DARTS, we set our model with 16 cells.
Reduction cells (stride=2) are the 1st, 5th, and 13th cells, and three FC layers are composed as in the small model.
For the other hyperparameters, we adopt the default setting in the cifar-10 example of DARTS~\cite{NASLiuSY19}.

The performance filter is placed before the original searching strategy in DARTS.
The channel number of convolution. layers is a hyperparameter in DARTS and it largely influences the computation cost.
With a predefined computation cost, the performance filter can estimate the channel number of convolution. layers for the formal searching process. 
The performance filter evaluates the forward pass of models by using PyTorch-OpCounter~\cite{pytorchOpCounter} and calculating the overall training cost by the aforementioned formula in Equation~\ref{eq:1}.
We set the candidate models with 4 TFLOPs (medium) and 16 TFLOPs (large) in the performance filter.
Eventually, the medium model has $101.6\times10^6$ parameters and the training process of a single sample requires about 4.15 TFLOPs, which largely exceeds that of the small model.
As for the large model, it has $374.2\times10^6$ parameters and takes 16.2 TFLOPs for training with a single sample.
Notably, we can store model architectures searched for using in other HPC systems.

\subsection{Parallelized Training}

Training a CNN model generally requires iterative traversals over the dataset to improve the accuracy of the model and make it finally converge.
We adopt the commonly-used data parallelism for the training process of SAIH-cosmo on the target HPC system.
Data parallelism allocates a training task on a minibatch of samples to an available computational devices (GPU or CPU core).
Each device keeps a copy of the target model and performs the training process independently on its own samples.
After finishing the forward and backward propagations, the weights are updated by performing allreduce operation on all devices.
We adopt NVIDIA NCCL as the collective data movement backend, which transmits data through NVLink and Infiniband.

Batch size, that determines the number of samples parallelly trained in a GPU, is critical for data parallelism. 
The larger batch size, the less number of allreduce operations are required for each epoch. 
This results in less communication overheads.
However, the batch size is limited by two factors, i.e., memory capacity and model convergence. 
In terms of memory capacity, increasing batch size leads to a proportional growth of memory capacity.
In order to maximize the performance of HPC system and investigate the bottleneck when it is fully loaded on GPU device, we tune the batch size for each model.
Thus, the batch sizes of the small model, medium model and large model in each GPU are set to 10, 4 and 1, respectively.
As for the convergence problem, it is mainly solved by optimizing machine learning algorithms, such as gradient-based optimization, e.g., momentum, Adagrad and Adam~\cite{AdamAlgorithm}.
Here we adopt Adam as our gradient optimizer.
Adam designs adaptive learning rate for accelerating stochastic optimization.
We set hyperparameters of Adam with $\beta_1=0.9$, $\beta_2=0.999$, $\hat{\epsilon}=10^{-8}$.
Then we further combine Adam with LARS~\cite{LARS}.
LARC adaptively adjusts the learning rate for each layer and it enables better convergence in extremely large batch size.

\section{Evaluation}
\label{eval}

\subsection{Experiment Setup}

SAIH-cosmo is implemented in PyTorch v1.5.0 and the \emph{DistributedDataParallel} class of PyTorch is used for data parallelism.
We perform both node scaling and data scaling executions on SAIH-cosmo with different models to understand 3D CNN training on the HPC system. 
Meanwhile, we further investigate their arithmetic intensity by the NVIDIA profiling tool (\emph{nvprof}) and mixed precision training by \emph{Nvidia apex}.
Overall, as shown in Table~\ref{tab:traces}, for each model, the node scaling training uses 1/32 of the complete dataset (101,088 samples) and runs on 1, 2, 4, 8, 16, and 32 nodes of the cluster ($n=1, 2, 4, 8, 16, 32$).
The data scaling training uses 4 nodes and the training dataset size scales from 1/64 to 1/1 of the complete dataset ($d=64, 32, 16, 8, 4, 2, 1$).
To illustrate the model convergence, we execute each training with 60 epochs.
Notably, it is optional for SAIH users to ignore the convergence and only focus on performance trends, which can largely reduce the evaluation costs, from days to a few hours.

\begin{table}[H]
	\centering
	\caption{Experiment Settings.}
	\label{tab:traces}
	\begin{tabular}{cccccc}
		\hline
		& \multicolumn{2}{c}{Node scaling}& & \multicolumn{2}{c}{Data scaling} \\ \cline{2-3}\cline{5-6}
		& Data               & Node       &      & Data               & Node    \\ \hline 
		Small Model & 1/32               & $n$         &     & $1/d$              & $4$     \\ 
		Medium Model & 1/32               & $n$        &      & $1/d$              & $4$     \\ 
		Large Model & 1/32               & $n$         &     & $1/d$              & $4$     \\ \hline
	\end{tabular}
\end{table}

\subsection{Model Accuracy}

\begin{figure*}
	\centering
	\subfigure[]{
		\includegraphics[width=0.29\linewidth]{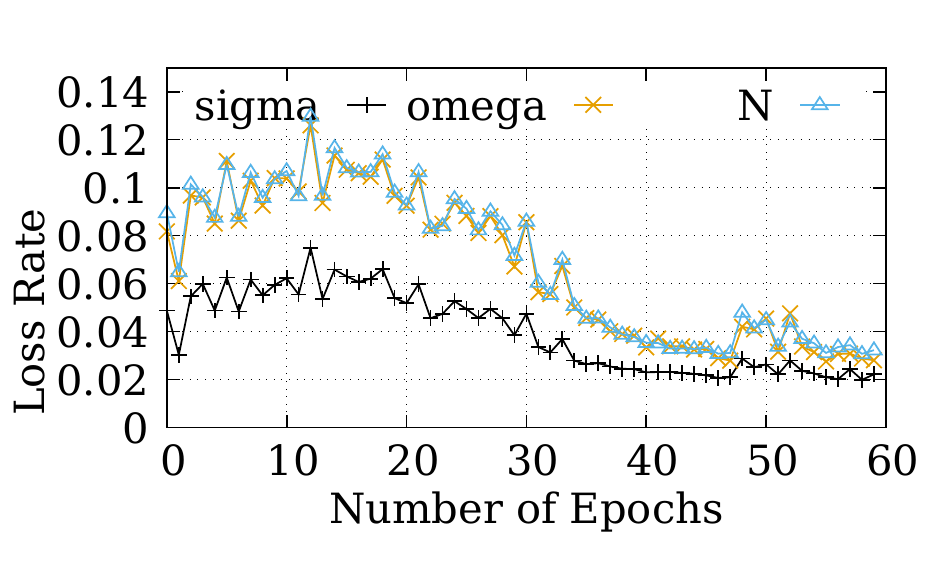}
		\label{fig:accuracy1-first}
	}
	\hfill
	\subfigure[]{
		\includegraphics[width=0.29\linewidth]{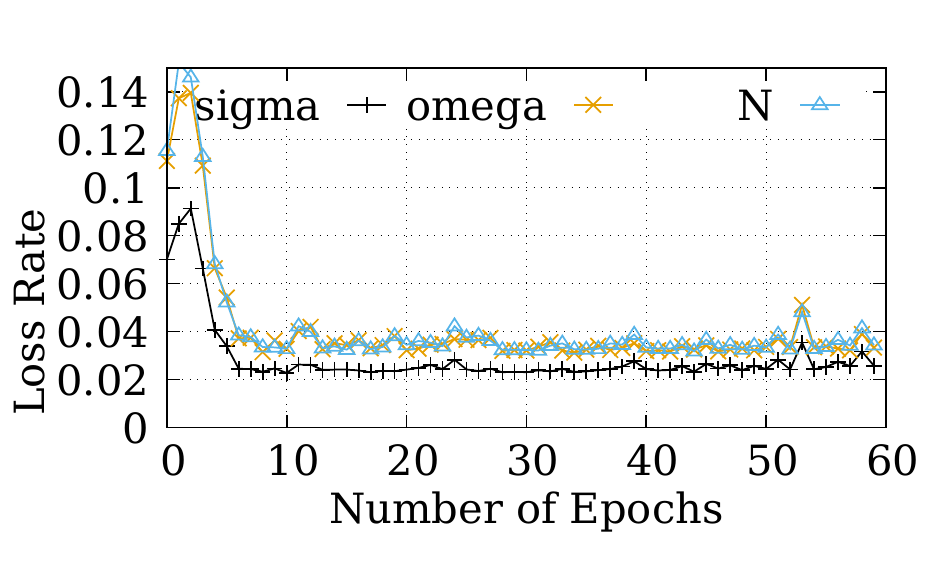}
		\label{fig:accuracy1-second}
	}
	\hfill
	\subfigure[]{
		\includegraphics[width=0.29\linewidth]{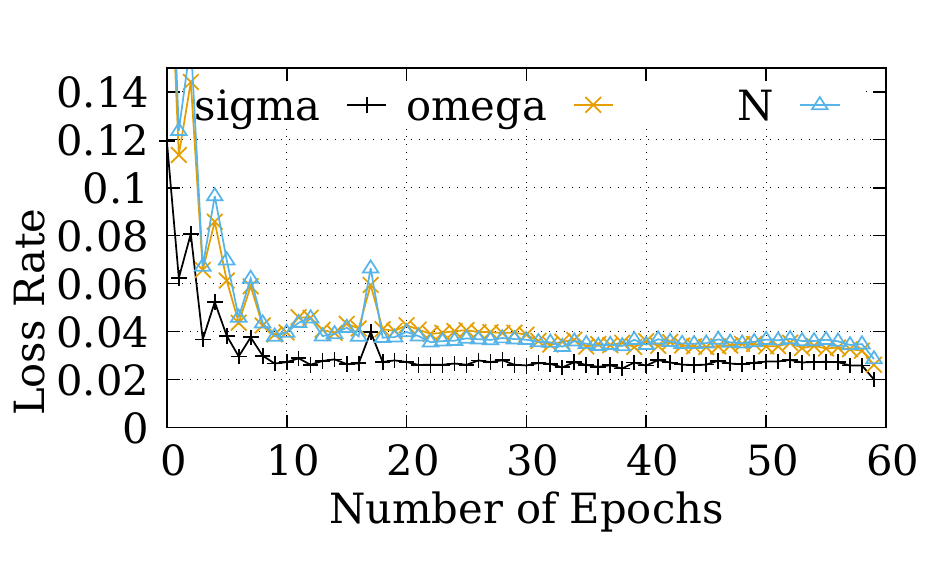}
		\label{fig:accuracy1-third}
	}
	
	\subfigure[]{
		\includegraphics[width=0.29\linewidth]{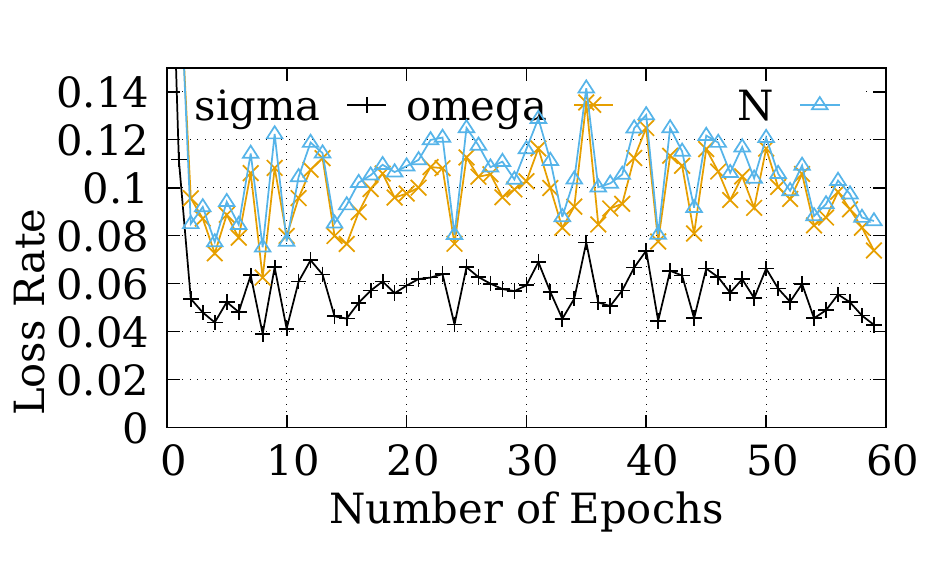}
		\label{fig:accuracy4-first}
	}
	\hfill
	\subfigure[]{
		\includegraphics[width=0.29\linewidth]{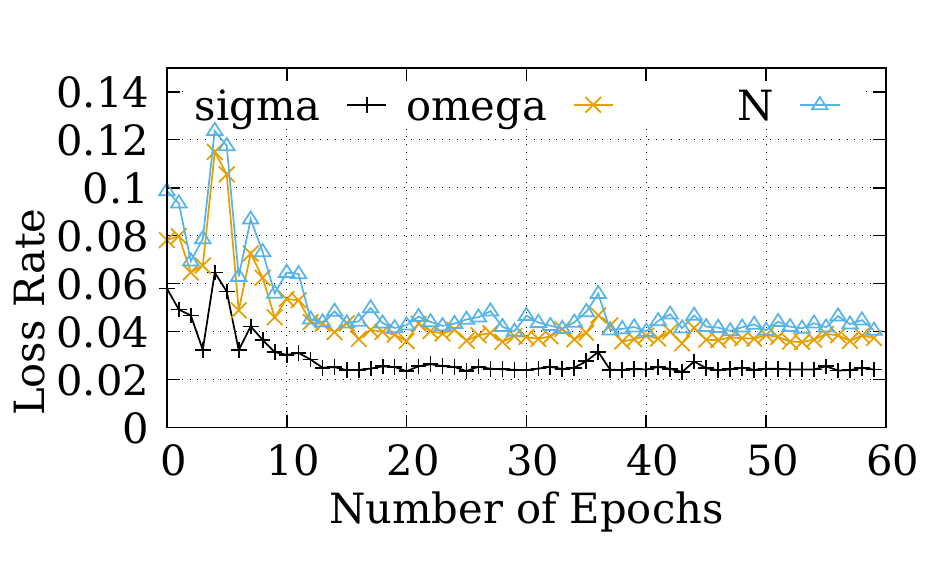}
		\label{fig:accuracy4-second}
	}
	\hfill
	\subfigure[]{
		\includegraphics[width=0.29\linewidth]{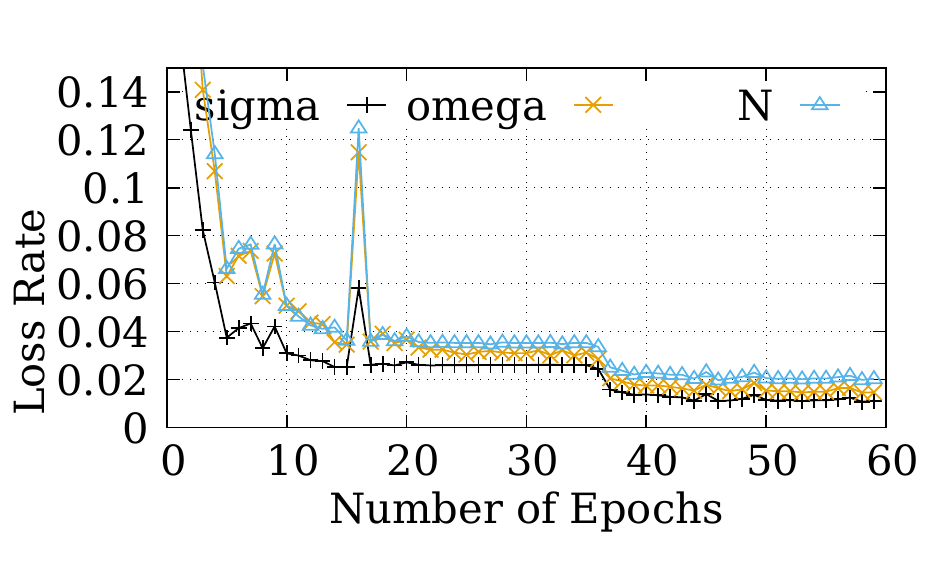}
		\label{fig:accuracy4-third}
	}
	\caption{Loss change in 1/32 dataset. (a) Small model in 1 node. (b) Medium model in 1 node. (c) Large model in 1 node. (d) Small model in 4 nodes. (e) Medium model in 4 nodes. (f) Large model in 4 nodes.}
	\label{fig:accuracy} 
\end{figure*}

Training an accurate DNN model is the fundamental requirement for an AI application.
The losses of three models show a rapid decreasing within the first epoch and then show different trends.
We demonstrate the losses after each epoch of three models in Fig.~\ref{fig:accuracy}, which shows that the larger model can generally provide better accuracy.  

Besides, we additionally find that the larger model in SAIH-cosmo has more stable convergence in multi-node training.
For the small model, the losses of three cosmological states decrease with significant fluctuations, and it fails to converge within 60 epochs.
In contrast, the medium and large models converge much more stable.
At the beginning of training phase, the losses of the medium model initially go up and then decrease rapidly, while those of the large model converges constantly and steadily.
Also, in the final epochs, the large model achieves with the best losses.

Although we apply LARC algorithm, 
increasing the number of GPUs to 16 (4 nodes) influences the convergence of the AI models.
As shown in Fig.~\ref{fig:accuracy4-first}, the small model displays wider fluctuations in losses, severely damaging the convergence, which is also in consistent with the results of cosmoflow~\cite{mathuriya2018cosmoflow}.
In terms of the medium model, it also shows more rough converging process compared with the 1-node execution, and slightly impairs the final accuracy.
On the contrary, distributed training contributes to positive impact on the large model.
Increasing the number of nodes makes the large model converges to a better result.

While we adopt the simplified NAS in SAIH-cosmo with a narrow searching space, our automatically generated models achieve significant improvement as the computation demand increases, which validates its applicability as well as the preservation of scientific meaning.

\textbf{Finding 1:} Overall, the models augmented by NAS lead to better accuracy and better convergence. Besides, with progressively increasing the sizes of models, we find that the convergences of larger models are more steady in multi-node training.


\begin{figure*}[!t]
	\centering
	\subfigure[]{
		\includegraphics[width=0.3\linewidth]{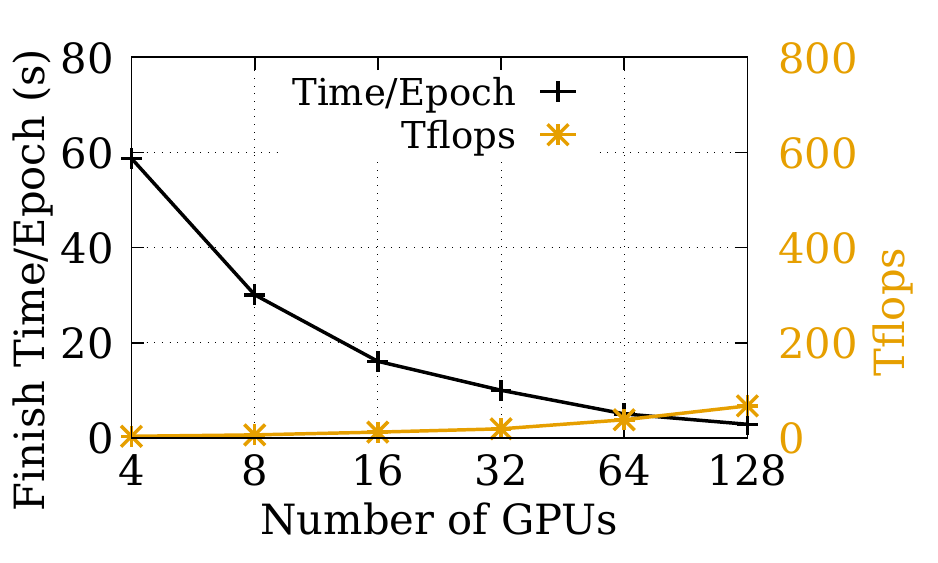}
		\label{fig:nodescaling-first}
	}
	\hfill
	\subfigure[]{
		\includegraphics[width=0.3\linewidth]{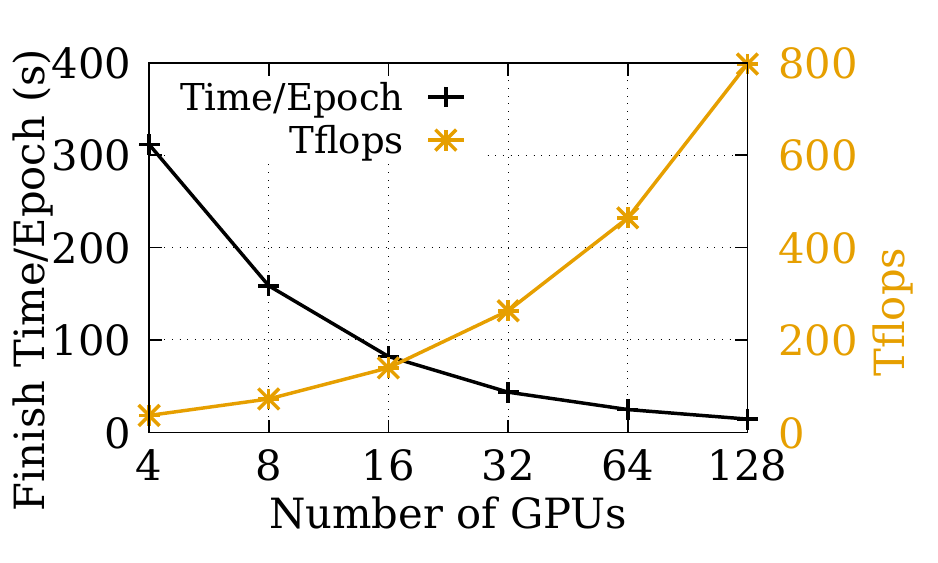}
		\label{fig:nodescaling-second}
	}
	\hfill
	\subfigure[]{
		\includegraphics[width=0.3\linewidth]{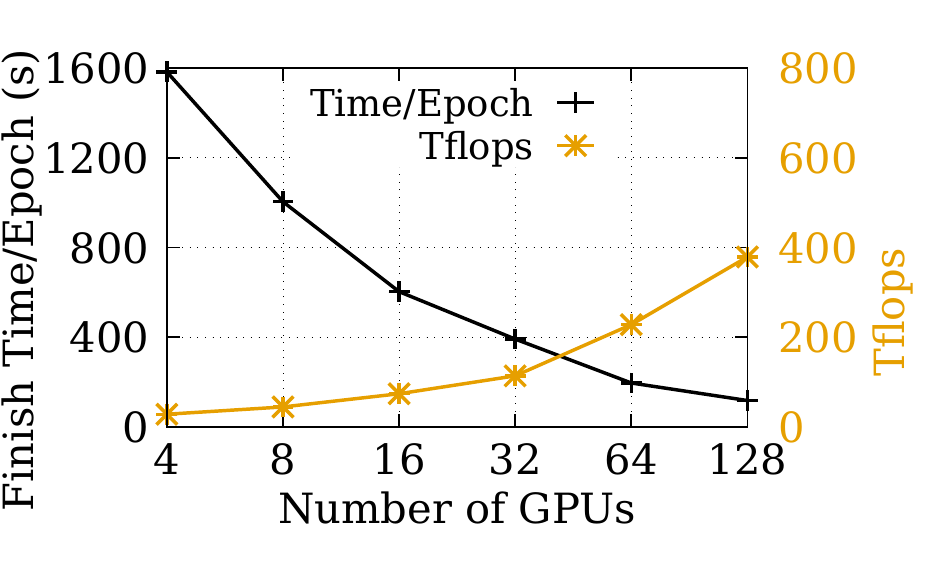}
		\label{fig:nodescaling-third}
	}
	
	\subfigure[]{
		\includegraphics[width=0.3\linewidth]{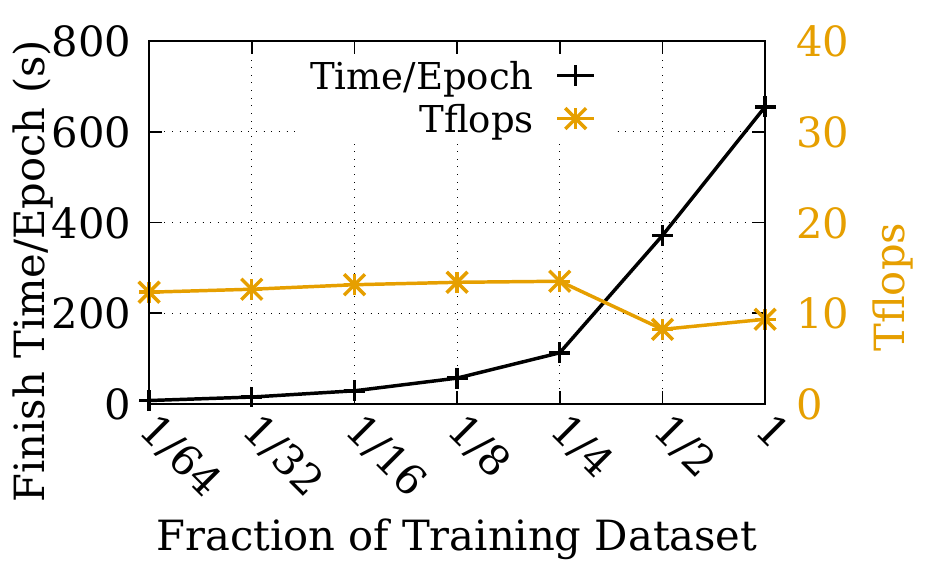}
		\label{fig:datascaling-first}
	}
	\hfill
	\subfigure[]{
		\includegraphics[width=0.3\linewidth]{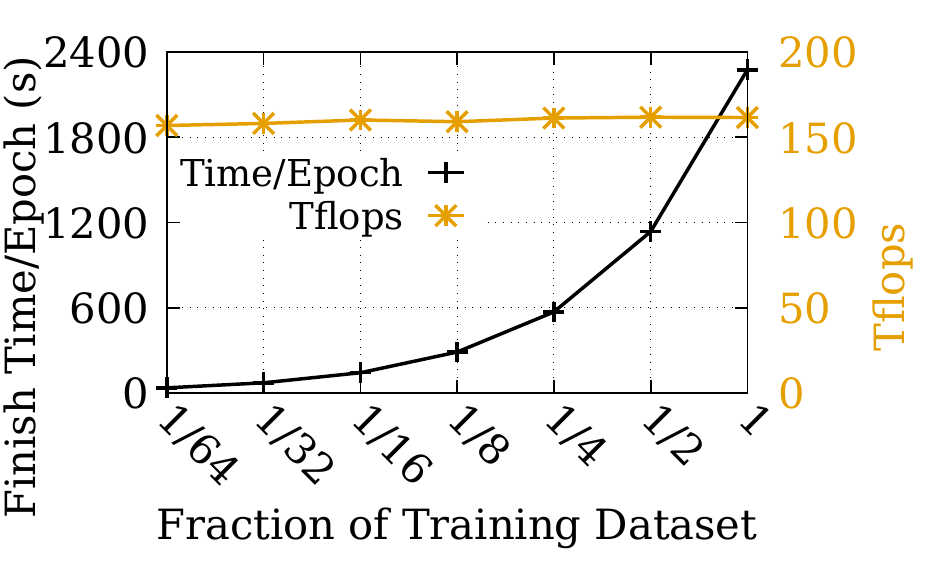}
		\label{fig:datascaling-second}
	}
	\hfill
	\subfigure[]{
		\includegraphics[width=0.3\linewidth]{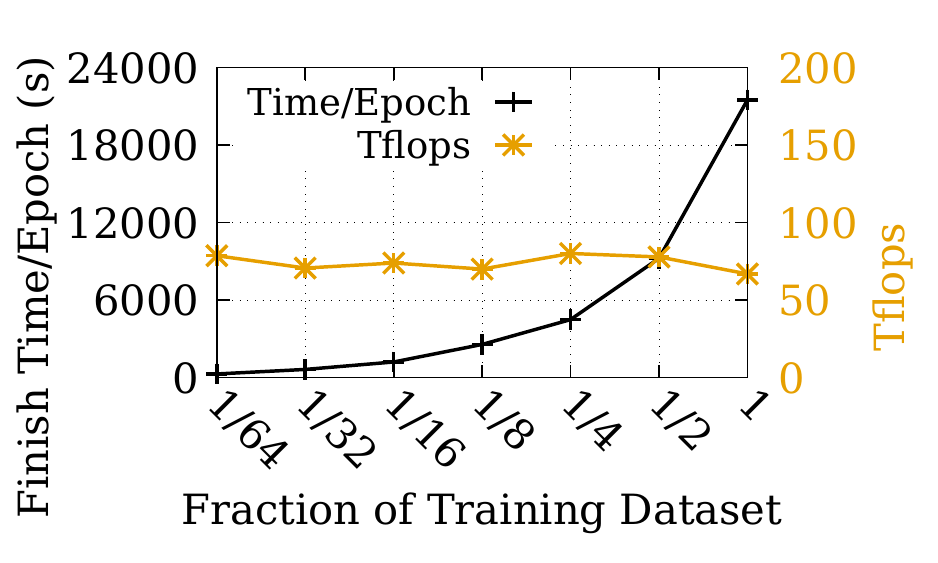}
		\label{fig:datascaling-third}
	}
	\caption{Performance evaluation with node and data scaling. (a) Small model in node scaling. (b) Medium model in node scaling. (c) Large model in node scaling. (d) Small model in data scaling. (e) Medium model in data scaling. (f) Large model in data scaling.}
	\label{fig:overallscaling} 
\end{figure*}

\subsection{Evaluation with GPU Node Scaling}

\begin{figure}[H]
	\centering
	\subfigure[]{
		\includegraphics[width=0.32\textwidth]{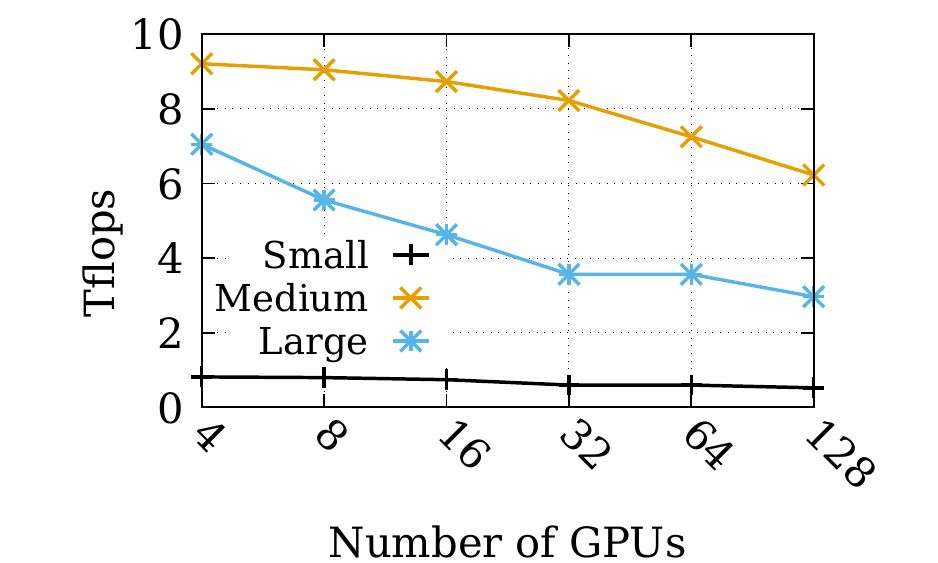}
		\label{fig:singleGPUnodescaling-first}
	}
	\hfill
	
	\subfigure[]{
		\includegraphics[width=0.3\textwidth]{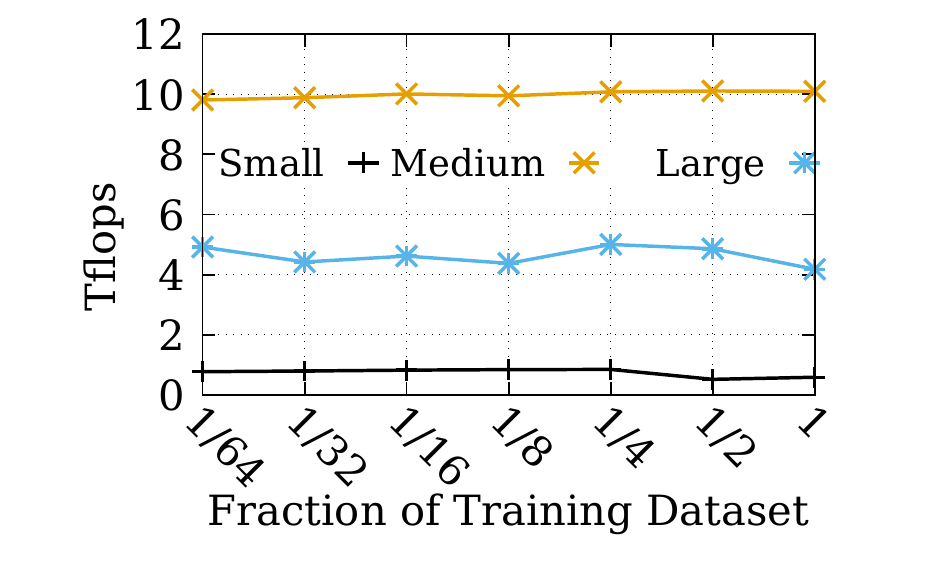}
		\label{fig:singleGPUdatascaling-second}
	}
	\caption{Average Tflops per GPU. (a) GPU Node Scaling. (b) Data Scaling.}
	\label{fig:singlescaling} 
\end{figure}

The evaluation with GPU node scaling follows the definition of strong scaling, which concerns the acceleration for a fixed problem size with respect to the number of GPU devices, and is limited by the fraction of serial part in a program that is not amenable to parallelization.
When fixing the data size (1/32 dataset) and scaling up the number of GPUs from 4 to 128, it displays different performance scalability for these models as shown in Fig.~\ref{fig:overallscaling}(a)(b)(c).
The average flops of a single GPU is shown in Fig.~\ref{fig:singleGPUnodescaling-first}.

We first discuss on the average performance of each GPU, which varies largely in different model sizes.
With 4 GPUs, the medium model shows the best performance at 9.21 Tflops for each GPU, achieving 58.66$\%$ peak performance of Tesla V100.
In comparison, the large model executes with lower average flops. 
The small model has only achieved about 800 Gflops on each GPU, which is much lower than the other two models.

As shown in Table~\ref{tab:arithintensity}, the arithmetic intensity of the small model is much lower than those of the other two models.
Here `Read and `Write' represent the number of memory read accesses and memory write accesses.
Also, we can notice that the computation demand of the small model is several orders of magnitude lower than those of the other two models.
To profile the memory usage of each model, we find that the gap of memory footprint between the small model and other two models is not as large as the computation demand.
In details, the weights and intermediate results of medium and large models are only 3.79$\times$ and 7.46$\times$ more than those of the small model. 
In comparison, the computation demands are 67.25$\times$ and 257.97$\times$ higher than those of the small model.
When we adjust the batch sizes of the three models to saturate the GPU memory, the small model cannot train with an extremely large batch size to make full use of all computing units, resulting in a much lower arithmetic intensity than the other models with more computation demands.
In this way, the small model is limited by GPU memory capacity and the small batch size cannot saturate the computing unit.
Besides, we also observe that the memory footprint mainly comes from intermediate results rather than weights in 3D CNN, where the memory footprint of intermediate result is about 30$\times$ larger than those of the weights in these three models.

\begin{table*} 
	\centering
	\caption{Arithmetic Intensity Analysis with 4 GPUs and 1/32 Dataset.}
	\label{tab:arithintensity}
	\begin{tabular}{c|c|c|c|c|c}
		\hline
		Model Size & Batch size  & FLOPs         & Read         & Write         & Intensity \\ \hline \hline
		S & 10          & $6.9\times 10^{10}$   & $1\times10^{7}$     & $7.53\times10^7$     & 808    \\ \hline
		M & 4           & $4.64\times10^{12}$  & $1.15\times10^{9}$  & $7.09\times10^{8}$   & 2500   \\ \hline
		L & 1           & $1.78\times10^{13}$  & $7.13\times10^{9}$  & $5.21\times10^{9}$   & 1442   \\ \hline
	\end{tabular}
\end{table*}

As for the large model,
although it has the largest GPU memory footprint and computation demand, the counter-intuitive observation is that it has lower arithmetic intensity than the medium model.
Table~\ref{tab:detailedarithintensity} and Table~\ref{tab:kernels} list the arithmetic intensity of top-10 kernels sorted by the number of memory access.
$\%$MemAcc is the percentage distribution of memory access of kernels.
'ID' is a unique ID for each kernel in CNN model, that is executed on GPU.
First, the types and percentages of kernels in the large model are quite different with those of the medium model.
In the large model, the kernel with 66.53$\%$ memory accesses holds only 516 arithmetic intensity, while the top-3 kernels in the medium model have 1.8$\times$ higher weighted aggregate arithmetic intensity than those of the large model, that is 2,183 compared to 1,210.
This is due to higher internal thread-level parallelism in the medium model than that in the large model.
In addition, the fact that the 3D feature map of the large model is bigger than that of the medium model results in more strided memory access, further decreasing the arithmetic intensity.

\textbf{Finding 2:}
For 3D CNN models, the ratio of memory footprint and computation amount is not stable but changes in a wide range.
Besides, the models with more layers and weights are not always leading to higher GPU performance and Tflops, which are instead significantly determined by the arithmetic intensity of AI models. 

Then we analyze the performance scalability of the distributed training.
When scaling the training to 128 Tesla V100 GPUs,
the small model achieves 112.3 aggregate Tflops; the medium model achieves up to 797.1 Tflops; and the large model achieves 379.2 Tflops.
From 4 GPUs to 128 GPUs, the speedups of these three models are 25.6$\times$, 22.2$\times$, and 13.45$\times$ respectively.
Since training procedure contains computation and communication in each iteration, the larger model has weaker scalability due to both the larger communication overhead of each iteration and the smaller batch size limited by the GPU memory capacity.
On the one hand, communications among GPUs adopt allreduce to collectively aggregate gradients with $O(parameters+nodes)$ time complexity.
The hyperparameter, \emph{batch size}, does not change the total number of gradients in the model to be communicated, since the gradients in a minibatch of samples will be aggregated locally within a GPU before performing inter-node communications.
On the other hand, the larger batch size can reduce the number of communication operations in one epoch, leading to relatively less communication overheads.
The 16 GB HBM of a single Tesla V100 GPU, which is larger than most of other GPUs' memory capacity, can only contain one input sample trained on large model, which occupies more than 13 GB memory. 
Thus, the larger model has smaller batch size, resulting in weaker parallelism.
Large batch size plays a key role in increasing parallel efficiency, which is also stressed in You et al.~\cite{LARS}.

\textbf{Finding 3:}
Besides saturating computation units by high parallelism, large batch size can further improve the performance scalability by reducing I/O frequency and overhead.

\begin{table}[H]
\centering
\caption{Profiling Results of CNN Kernels.}
\label{tab:detailedarithintensity}
\resizebox{0.95\columnwidth}{!}{
\begin{tabular}{ccccccc}
\hline
 \multicolumn{3}{c}{Medium Model}& & \multicolumn{3}{c}{Large Model} \\ \cline{1-3} \cline{5-7} 
ID& \%MemAcc     & Intensity    && ID    & \%MemAcc    & Intensity    \\ \hline
0 & 26.76\%      & \textbf{922} && 0     & 66.52\%           & \textbf{516} \\ 
1 & 26.31\%      & \textbf{5201}&& 1     & 13.57\%           & \textbf{6391}\\ 
2 & 15.06\%      & \textbf{3777}&& 5     & 2.18\%            & \textbf{5}   \\ 
3 & 5.68\%       & 19           && 10    & 1.57\%            & 19           \\ 
4 & 3.41\%       & 1            && 11    & 1.46\%            & 4            \\ 
5 & 3.40\%       & 36           && 3     & 1.46\%            & 4            \\ 
6 & 2.90\%       & 2            && 7     & 1.46\%            & 4            \\ 
7 & 2.90\%       & 8            && 6     & 1.10\%            & 8            \\ 
8 & 1.75\%       & 0            && 12    & 1.02\%            & 1            \\ 
9 & 1.15\%       & 12           && 13    & 0.94\%            & 36           \\ \hline
\end{tabular}
}
\vspace{-10pt}
\end{table}

\begin{table}[H]
	\footnotesize
	\centering
	\caption{ Kernels in Cosmological CNN Model.}
	\label{tab:kernels}
	\resizebox{0.45\textwidth}{!}{
		\begin{tabular}{cc}		
			\hline
			ID & Kernel \\ \hline
			0 & $scudnn\_128x64\_stridedB\_splitK\_small\_nn\_v1$ \\
			1 & $scudnn\_128x64\_stridedB\_splitK\_medium\_nn\_v1$ \\
			2 & $scudnn\_128x128\_stridedB\_splitK\_small\_nn\_v1$ \\
			3 & $implicit\_convolveNd\_sgemm (int 1024)$ \\
			4 & $convolveNd\_wgrad\_engine$ \\
			5 & $implicit\_convolveNd\_sgemm (int 512)$ \\
			6 & $bn\_bw\_1C11\_kernel_new$ \\
			7 & $bn\_fw\_tr\_1C11\_kernel\_NCHW$ \\
			8 & $setOutputKernel$ \\
			9 & $vectorized\_elementwise\_kernelGLOBAL\_\_N\_\_57\_tmpxft$ \\
			10 & $convolveNd\_wgrad\_engine (int=8)$ \\
			11 & $convolveNd\_wgrad\_engine (int=7)$ \\
			12 & $vectorized\_elementwise\_kerneladd\_kernel\_cuda$ \\
			13 & $vectorized\_elementwise\_kernelgpu\_kernel\_with\_scalars$ \\
			\hline
			\end{tabular}}
\end{table}

\subsection{Evaluation with Data Scaling}


In this subsection, we further investigate the data scaling, which fixes the number of nodes to 4 and scale the training dataset from the fraction 1/64 to 1/1 of the complete dataset and the performance is illustrated in Fig.~\ref{fig:overallscaling}(d)(e)(f).
The average flops per GPU is in Fig.~\ref{fig:singleGPUdatascaling-second}.

Overall, in data scaling, there are two major factors that affect the aggregate performance, data loading in memory hierarchy and communication for allreduce.
For communication, since the number of communication-intensive operation, allreduce, changes proportionally to the computation, it almost has no impact on the ration of communication time and the overall evaluation time. 
Thus, we here mainly focus on data loading.
In terms of the medium model and the large model, their performances fluctuate in a limited range, and that of the large model is wider.
As for the small model, when the data volume is 1/64, 1/32, 1/16, 1/8, and 1/4, the average flops is very stable at around 800 Gflops, while there is a decrease to about 520 Gflops with the 1/2 and 1/1 datasets, and this observation is more obvious in Fig.~\ref{fig:datascaling-first}. 
Investigating the breakdown, for all of the three models, we observe that there is a long data loading time at the beginning of each epoch, and unexpected data loading time when training with the 1/1 or 1/2 datasets.
As the small model has the shorter computation time and larger fraction of data loading time, we can observe that the overall performance decrease is more obvious.

The data loader in PyTorch endeavors to load all training data into GPU node memory.
At the beginning of each epoch, the execution of data loading results in a large amount of idle time for computing units, as well as heavily stressing the shared file system.
Zooming into the unexpected data loading time, the data loader in PyTorch has a prefetch strategy from node memory to GPU memory, while it lacks an efficient data prefetch strategy from file system to node memory.
The entire dataset takes 1.6 TB storage, while each node of the cluster has only 256 GB memory capacity.
In this way, 4 nodes fail to load the entire 1/1 and 1/2 dataset into memory.
It has to wait for data in the swap area to be loaded before continuing training procedure, leading to unexpected data swapping overhead.
For example, the percentage of data loading time in small model increases from 2.83$\%$  to about 20$\%$ as the dataset scales from 1/4 to 1/1, resulting in 31.2$\%$ reduction in overall flops.

\textbf{Finding 4:} 
Different from traditional HPC applications, AI training demands I/O throughout during the whole training procedure for accessing input training data and intermediate data. Thus, a local storage system on each compute node is recommended for staging the training and intermediate data.

\textbf{Finding 5:}
Large dataset with frequent data accesses can make memory hierarchy become a bottleneck. 
Thus, with scaling scientific dataset, SAIH-cosmo probes the performance capability of memory hierarchy, which is critical to the evaluation of the overall AI performance on HPC systems.

\begin{table}[H]
\centering
\caption{Execution Time in 1/32 Dataset with Different Precision Levels.}
\label{tab:apex}
\resizebox{0.95\columnwidth}{!}{
\begin{tabular}{ccccc}
\hline
\multirow{2}{*}{    Precision level }   &\multirow{2}{*}{Description}  & \multicolumn{3}{c}{Execution time (s)} \\ \cline{3-5} 
               &                   & Small & Medium & Large   \\ \hline 
Apex[O0]            & pure FP32         & 18.71 & 364.06 & 2497.2  \\
Apex[O1]            & mixed precision   & 19.46 & 205.38 & 1867.9  \\ \hline
\end{tabular}
}
\vspace{-5pt}
\end{table}

\subsection{Mixed Precision Results}
Theoretically, Tesla V100 claims 15.7 Tflops of single-precision (FP32) performance and 125 Tflops half-precision (FP16) performance by utilizing its novel tensor core architecture.
We evaluate the performance of SAIH-cosmo under the mixed precision setting by employing NVIDIA \emph{apex} with optimization level \emph{O1}.
However, as shown in Table~\ref{tab:apex}, the mixed precision evaluations, i.e., \emph{Apex[O1]}, can only achieve up to 1.76$\times$ speedup for the three models compared with the FP32 execution, i.e., \emph{Apex[O0]}. This is significantly lower than the theoretical peak performance.

We adopt \emph{nvprof} with turning on the $tensor\_precision\_fu\_utilization$ metrics to diagnose the counter-intuitive performance gap.
While according to the NVIDIA official document, only cudnn newer than version 8.0.2 can be compatible with PyTorch and enable it to utilize tensor core architecture for training 3D convolution with mixed precision.
Our results show that the tensor core is under a very low utilization only contributing around 4.4$\%$ of the entire execution, which results in an extremely limited performance gain on SAIH-cosmo.
Also, through profiling on individual kernels, we find that cudnn is inefficient in some outlier mixed-precision 3D convolutions, e.g., the situation with a small number of channels requires high memory access overhead, and it is easy to improve.
To further take advantage of the tensor core architecture, the instruction-level restrictions under high-level AI frameworks and convolution libraries, e.g., the input size and memory layout~\cite{yan2020demystifying}, have to be resolved, in particular for performing 3D convolutions, which are not optimized as well as the 2D convolutions commonly used in hotspot AI research areas.

 \textbf{Finding 6:} AI related libraries, like cudnn and cublas, are released with the capability of supporting low-precision and mixed-precision training operations on novel architectures, e.g., tensor core.
 However, without customized by AI system experts, the mixed precision is not well prepared in a plug-and-play manner for supporting accelerating common AI operations on AI frameworks, e.g., conv3d in PyTorch.

\subsection{Comparison and Summary}
\begin{table*}
\caption{SAIH and Representative Benchmarks.}
\label{tab:bmcomparsion}
\resizebox{0.95\textwidth}{!}{
\begin{tabular}{lllll}
\hline
\textbf{Name}       & \textbf{Positioning}                                                                                 & \textbf{Metrics}                                                                 & \textbf{Application Domain}                                                        & \textbf{Targets}                                                                                                                  \\ \hline
HPL-AI     & \begin{tabular}[c]{@{}l@{}}Benchmark\end{tabular}       & \begin{tabular}[c]{@{}l@{}}Flops,\\ Flops/Watt\end{tabular}             & Linear equations                                                         & \begin{tabular}[c]{@{}l@{}}Measures mixed-precision performance \\ of HPC systems.\end{tabular}                           \\ \hline
Deep500    & \begin{tabular}[c]{@{}l@{}}Benchmark \end{tabular}       & \begin{tabular}[c]{@{}l@{}}Throughput,\\ Time to solution\end{tabular}  & \begin{tabular}[c]{@{}l@{}} Any AI applications\\ in theory\end{tabular}                                                       & Helps evaluate framework implementations. \\ \hline
MLPerf-HPC & \begin{tabular}[c]{@{}l@{}}Benchmark \end{tabular}       & Time to solution                                                        & \begin{tabular}[c]{@{}l@{}}Cosmology and \\ weather analysis\end{tabular} & Speeds up the training time.                                                                                           \\ \hline
HPC-AI500  & \begin{tabular}[c]{@{}l@{}}Benchmark\end{tabular}       & Valid FLOPS                                                                             & \begin{tabular}[c]{@{}l@{}}Imagenet and \\ weather analysis\end{tabular} & Speeds up the training time.                                                                                          \\ \hline
SAIH       & \begin{tabular}[c]{@{}l@{}}Evaluation Method\\ (mainly vertical comparison)\end{tabular} & \begin{tabular}[c]{@{}l@{}}Flops as scaling,\\ Scalability\end{tabular} & Cosmology                                                                 & \begin{tabular}[c]{@{}l@{}}Understands AI Performance Trend and guide \\ the design for emerging HPC system.\end{tabular} \\ \hline
\end{tabular}
}
\end{table*}

This subsection is going to compare SAIH with other methods and illustrate our contributions.
Notably, SAIH is an evaluation methodology rather than an AI benchmark with defined AI model and dataset, e.g., HPL-AI and HPC-AI500.
It is different from existing benchmarking methods on evaluation target (AI performance trend via both strong scaling and weak scaling evaluations), evaluation settings (varying model sizes and dataset sizes) and evaluation testcases (cosmological AI applications). 
Specifically, in our testcase, we adopt unique evaluation settings such as both strong scaling and weak scaling evaluations with varying model sizes and dataset sizes. 
Thus, it could not directly compare our method to existing benchmarks, and we make a  qualitative comparison between SAIH and 4 representative benchmarks, i.e., HPL-AI~\cite{haidar2018harnessing}, Deep500~\cite{Deep500}, MLPerf-HPC~\cite{mlperfhpc} and HPC-AI500~\cite{hpcai500}, to illustrate that our SAIH is effective and can bring new insights for next-generation HPC systems.

At first, SAIH has a different positioning from the benchmark works.
Benchmark works mainly focus on the AI performance of the same job in different infrastructures and rank different infrastructures, while SAIH aims at demonstrating the performance changes when AI applications scale up and understanding the requirements of AI applications for subsystems, e.g., I/O, compute unit, and communicate.
In detail, HPL-AI can represent the traditional HPC benchmarks.
It measures and compares the mixed-precision performance of HPC systems by solving mixed-precision linear equations.
Notably, HPL-AI provides the scalability as our SAIH and can derive the maximum performance of HPC systems.
However, it mainly investigates the computation of an HPC system and can hardly represent the practical AI performance of an HPC system.
In comparison, the data and model augment methods in our SAIH keep the scientific meaning when scaling AI applications, making our evaluation with practical significance.

Deep500 is a modular benchmark infrastructure can be integrated into most evaluation methods.
It characterizes many fine-grained metrics, e.g., utilization of computing devices, collective communication, and IO, into modules and can be used to help evaluate different framework implementations and multiple levels of operators.
In fact, the implementation of Deep500 is difficult to be integrated into large-scale AI applications and our SAIH includes many metrics defined in Deep500.

MLPerf-HPC is an emerging benchmark focusing on the intersection of scientific AI applications and HPC systems.
MLPerf-HPC rewrites the cosmoflow (as our SAIH) and a weather analysis AI application.
Compared with SAIH, it mainly adopts static dataset and model architecture and takes the traditional time-to-solution metric as the main criterion for ranking HPC systems, lacking of scalability.
Compared with SAIH, it lacks the weak scaling evaluation, which is important in the HPC filed.
Thus, it cannot investigate the performance changes dynamically and help understand the AI performance trends as our SAIH.

At last, for HPC-AI 500, in order to further assure equivalence, it presents a new metric named Valid FLOPS, which imposes a penalty on failing to achieve a target training quality. It selects the image classification and extreme weather analysis as tasks. Besides, its core idea is similar to MLPerf-HPC.
The findings of our SAIH are different from those of MLPerf-HPC and HPC-AI500, and can illustrate the AI capability and performance trend of HPC systems in both strong and weak scaling evaluation.

\begin{table*}	\footnotesize
	\centering
	\caption{Performance and Qualitative Evaluation Summary of SAIH-cosmo.}
	\label{tab:cosmoflow}
	\resizebox{0.65\textwidth}{!}{
		\begin{tabular}{l|l|l|l}		
			\hline
			\textbf{Metrics} & \textbf{Value} & 	\textbf{Metrics} & \multicolumn{1}{c}{\textbf{Value}} \\ 
			\Xhline{3\arrayrulewidth}
			\textbf{Domain} & Cosmology & \textbf{Data augment} & Simulation  \\
			\hline
			\textbf{Model augment} & NAS &\textbf{DNN model} &3D CNN \\
			\hline
			\textbf{Data format} &  3D particle  &	\textbf{AI framework}&  PyTorch  \\
			\hline

			\textbf{Dataset size} & \multicolumn{3}{c}{0.025 - 1.6 TB} \\
			\hline
			\textbf{HPC system} & \multicolumn{3}{c}{\begin{tabular}[c]{@{}c@{}}GPU cluster with 32 nodes, 4 Tesla V100 GPUs/node \end{tabular} } \\
			\hline
			\textbf{FLOPs of AI model } &\multicolumn{3}{c}{ 69GFLOPs - 16.2TFLOPs} \\
			\hline
			\textbf{flops (\% of theoretical flops)} & \multicolumn{3}{c}{69.6 (5.2\%) - 797.1 Tflops (59.6\%) } \\
			\hline
			\textbf{Singe GPU performance} &\multicolumn{3}{c}{ 0.51 - 9.21Tflops} \\
			\hline
			\textbf{Speedup on strong scaling} & \multicolumn{3}{c}{13.45$\times$ - 25.6$\times$} \\
			\hline
			\textbf{Arithmetic Intensity} & \multicolumn{3}{c}{838 - 2501} \\
			\hline
			\textbf{Accuracy / Loss} & \multicolumn{3}{c}{0.02475 - 0.06754} \\
			\hline
		\end{tabular}}
\end{table*}

Through the evaluation and comparison, we notice that the main idea of SAIH does not overlap with existing methods.
SAIH provides a feasible solution to make AI application evaluation scalable and can investigate many novel findings.
Particularly, with the SAIH-cosmo, we evaluate the AI performance for an important type of model, 3D CNN, on a heterogeneous HPC cluster.
The evaluation with scaling problem size helps us diagnose some emerging system performance bottlenecks and investigate the potential AI performance of the HPC system.
Moreover, we illustrate performance details of SAIH-cosmo and quantify its evaluation on the HPC system in Table~\ref{tab:cosmoflow}.

\section{Conclusion and Future Work}
\label{conclusion}

In this paper, a novel scalable methodology, SAIH, is proposed for better understanding AI performance trend of HPC systems.
Based on the methodology, we implement a testcase SAIH-cosmo.
SAIH-cosmo is transformed from a cosmological AI application and it is with data and computation scalability.
Through evaluating SAIH-cosmo on a real HPC system, we successfully diagnose many new insight findings as data and computation scaling large.
For example, we find that the convergences of larger models are more steady in multi-node training, which cannot be noticed if there is not a range of incremental models.
In this way, with SAIH, we can not only diagnose existing system performance bottlenecks, but also guide emerging HPC systems towards better AI support.
In future work, we plan to enhance SAIH in the following aspects:  1) building new representative SAIH instances in various scientific domains, e.g., RNN application in genome analysis, 2) creating fine-grained profiling components so as to identify performance bottlenecks on HPC systems.

\end{document}